\documentclass[%
 twocolumn,
 floatfix,
 amsmath,amssymb,
 aps,
pre,
]{revtex4-2}

\usepackage{graphicx}
\usepackage{dcolumn}
\usepackage{multirow}
\usepackage{bm}

\usepackage{float}
\restylefloat{table}

\def\<{\langle}
\def\>{\rangle}

\def\g2{g^{(2)}}
\DeclareMathAlphabet\mathbfcal{OMS}{cmsy}{b}{n}

\newcommand{\bra}{\langle}
\newcommand{\ket}{\rangle}

\newcommand{\non}{\nonumber}
\newcommand{\be}{\begin{equation}}
\newcommand{\ee}{\end{equation}}
\newcommand{\bea}{\begin{eqnarray}}
\newcommand{\eea}{\end{eqnarray}}

\begin{document}

\preprint{}

\title{Isothermal and adiabatic elastic constants from virial fluctuations}

\author{Andrey Pereverzev}
 \email{pereverzeva@missouri.edu}
\affiliation{%
 Department of Chemistry, University of Missouri,\\ Columbia, Missouri 65211-7600, USA
}%


\date{\today}

\begin{abstract}
We derive expressions for classical isothermal and adiabatic elastic constants
for periodic systems with the boundary contributions included explicitly.  The potential-dependent part of these expressions is written in terms of 
potential energies of atomic groups that make up the total potential energy. It is shown that in the thermodynamic limit the Born term, which depends on the second derivatives of potential energy, can be expressed exactly in terms of    
equilibrium averages that involve two types of atomic-group virials. As a result 
the new form of the Born term involves only first derivatives of either atomic-group or total potential energies.
The derived elastic constants expressions involving the two forms of the Born terms are tested and compared using molecular dynamics
simulations of crystalline argon and silicon. For both materials the elastic constants obtained using the two forms of the Born term are in good agreement. In particular, the new form of the Born term converges to the same value as the original Born term but at a slower rate. The results for silicon also agree well with the results from the previous molecular dynamics studies.
\end{abstract}

\maketitle

\section{Introduction}
Second order elastic constants describe the response of solid material stress tensor to 
imposed strain and, as such, they are important for characterizing materials properties under elastic deformation.
The microscopic theory of elastic constants was first developed by Born (see Ref. \cite{Born} and references therein) for the case of $T=0$. The theory was later extended to finite temperatures  \cite{Hoover1,Parrinello81, Parrinello82,Sprik,Ray88}.

The computational approaches for calculating elastic constants can be grouped into two categories:
equilibrium fluctuation methods and the direct method.

The equilibrium fluctuation methods can be further subdivided into the strain-fluctuation and the stress-fluctuation approaches.
 In the strain-fluctuation approach, the compliance tensor
is calculated from the strain fluctuations in the isothermal-isostress ensemble \cite{Parrinello81, Parrinello82,Sprik,Ray88}. The compliance tensor is then inverted to obtain elastic constants. 
The main disadvantage of this method is that it shows slow convergence and, as a result,  requires very long simulation trajectories.

In the stress-fluctuation method \cite{Hoover1,RR84,RR85,Ray88,Lutsko}, the elastic constants are calculated from the canonical or microcanonical ensemble averages and require evaluation of the so-called Born, kinetic, and stress fluctuations terms. 
The disadvantage of the strain-fluctuation approach is the necessity to calculate the Born term which involves the second derivatives of potential energy and can be quite
complicated. The complexity of the Born term calculation for arbitrary potentials 
makes it difficult to implement the strain-fluctuation approach in standard software
packages without great effort and risk of introducing undetectable
errors. Recently, a capability to numerically calculate the original form of 
the Born
term  for all potentials  was added to LAMMPS simulation package \cite{Plimpton}
based on the approach of Zhen and Chu \cite{ZhenChu}.

Numerical tests comparing the stress- and strain-fluctuation methods show that the stress-fluctuation approach provides faster convergence of the elastic constants compared to the strain-fluctuation approach \cite{Gao,Clavier}.

Because of the disadvantages of the stress- and strain-fluctuation methods 
just discussed,  the elastic constants are often calculated using the so-called direct method, in which finite
strains are applied to the system and the resulting changes in stress tensor are computed \cite{Hooks}.  This approach is fairly straightforward to implement but it is computationally demanding, especially for low-symmetry crystals.

Note that the formal elastic constants expressions in the case of the stress-fluctuation method are derived for infinite systems. Numerical simulations are  performed with finite systems, typically under periodic boundary conditions. The effect of periodic boundaries and how they are treated in 
simulations are rarely discussed explicitly  \cite{Hoover1,Lutsko}.

In this work we revisit elastic constants equations for the stress-fluctuation approach.
Our goal is 
two-fold. First, we re-derive expressions for isothermal and adiabatic elastic
constants in the form that explicitly includes the effect of periodic boundary conditions. Second, we show that the Born term can be written in a form that
involves only first derivatives of the potential energy. Apart from fundamental theoretical interest,  the new expression for the Born term can be used as the foundation of a new method of numerical calculation of isothertmal and adiabatic elastic constants.

\section{Isothermal and adiabatic elastic constants under periodic boundary conditions}
\subsection{The Hamiltonian of a periodic system} \label{one}
A conventional way to treat periodic boundary conditions is to assume that the system of interest which is placed in the central parallelepiped (or triclinic) box 
 is surrounded by an infinite number of translated identical image boxes that fill up the space  \cite{Erpenbeck,Todd,Bekker,Thompson}.
It is convenient to specify these boxes with 
vector ${\bf n}$ given by
\be
{\bf n}=\xi_a{\bf a}+\xi_b{\bf b}+\xi_c{\bf c}, \label{vector}
\ee
where ${\bf a}$, ${\bf b}$, and ${\bf c}$ are the vectors specifying the three edges of the central box and 
 $\xi_a$, $\xi_b$, and $\xi_c$ can take any integer values. The central box corresponds to ${\bf n}=0$. 
Then, the classical Hamiltonian of a system of $N$ atoms subject to periodic boundary conditions can be written as  
\be
H=H_0+ U\left(\{{\bf x}_{{\bf n }i}\}\right). \label{Hamiltonian}
\ee
Here 
\be
H_0=\sum_{i}^N\sum_{\alpha}\frac{p^2_{i\alpha}}{2m_i}
\ee
is the kinetic energy, in which $p_{i\alpha}$ is the $\alpha$th component of the momentum of atom $i$ with mass $m_i$ and $U\left(\{{\bf x}_{{\bf n }i}\}\right)$ is the system potential energy, which
depends on the set of atomic coordinates $\{{\bf x}_{{\bf n }i}\}$ both for the central box and the image boxes. These coordinates are given by
\be
{\bf x}_{i{\bf n }}={\bf x}_{i}+{\bf n },
\ee
where ${\bf x}_{i}$ is the coordinate vector of atom $i$ in the central box. For example, in the case of pair interactions, the potential energy can be written as  \cite{Erpenbeck, Thompson}
\be
U\left(\{{\bf x}_{{\bf n }i}\}\right)=\sum_{\bf n}\sum_{\substack{i=1\\j>i}}^{N} u(|{\bf x}_{i}-
{\bf x}_{j{\bf n }}|).
\ee
Here the sum over ${\bf n }$ extends over all image boxes. However, for potentials with a
cutoff and under the minimum image convention only the central box and its 26 
nearest-neighbor image boxes contribute to the sum over ${\bf n }$.
 
 An alternative but equivalent form of the potential energy
 for a periodic system
 that is useful for numerical applications can be 
 written in terms of the atomic-group potential energies as was done in Ref. \cite{Thompson}.
 The system potential energy is usually given 
 by a sum of two-atom, three-atom, four-atom or, in general, few-atom interaction terms. Each of these few-atom terms is commonly referred to as a group \cite{Thompson}
 and can be specified with an index $k$. The number of groups, the number of atoms in each group, and the number of groups in which an atom participates depends on the total potential energy but otherwise are completely arbitrary.
The potential energy of the group $k$ is denoted
by $u_k(\{{\bf x}_i^k\})$, where $\{{\bf x}_i^{k}\}$ is the set of coordinates for
group $k$ with the index $i$ running from 1 to $N_k$, the total number of atom in the group $k$. (Note that the same ${\bf x}_i$ for atom $i$ appears as ${\bf x}_i^{k}$ with a different $k$ for every group in which atom $i$ participates.)
The Hamiltonian of a periodic system in the atomic-group form can be written as 
\be
H=H_0+\sum_{k}u_{k}(\{{\bf x}_i^{k}\}), \label{Ham2}
\ee
where the sum over $k$ runs over all groups associated with the central box.
Thompson et al. \cite{Thompson} discuss in detail how such groups are defined
and selected.
Both  Eq. (\ref{Hamiltonian}) and Eq. (\ref{Ham2}) forms of the Hamiltonian are used below to derive
expressions for the elastic constants.
\subsection{The choice of strain} 
Elastic constants describe system response to the imposed strain and, in general,
depend on a particular form of strain used.
Consider a parallelepiped-shaped sample of solid material specified by three basis vectors starting from the common vertex.
Any homogeneous transformation of this sample can be represented by a real $3\times3$ matrix {\bf M} that transforms vectors
${\bf x}$ in the unstrained system to vectors ${\bf x}'$ in the strained system 
\be
 {\bf x}'={\bf M}{\bf x}.
\ee
The matrix ${\bf M}$ has nine independent components  of which six represent pure strain
and three represent rotation or reflection.
A convenient way to separate strain from rotation or reflection is to use the polar decomposition of the matrix ${\bf M}$ wherein ${\bf M}$  is uniquely written as  \cite{KornKorn}
\be
{\bf M}={\bf O}{\bf S}, \label{polar}
\ee
where ${\bf O}$ is an orthogonal matrix and ${\bf S}$ is a symmetric matrix.
Thus, the action of matrix ${\bf M}$ can be viewed as inducing a pure strain with the matrix ${\bf S}$ followed by a pure rotation or reflection, given by  the matrix ${\bf O}$, in which the strain state remains unchanged. Clearly, if ${\bf M}$ is symmetric then ${\bf M}={\bf S}$ and
the matrix ${\bf M}$ itself induces only pure strain with no rotation or reflection.
The difference between the matrix ${\bf S}$ and the identity matrix ${\bf I}$

\be
\bm{\mathcal{E}}={\bf S}-{\bf I} \label{E1}
\ee
arguably represents the
simplest definition of strain in that it is given by a linear deviation of the matrix ${\bf S}$ from the identity matrix.
From Eq. (\ref{E1})
the matrix {\bf S} is written in terms of $\bm{\mathcal{E}}$ as
\be
{\bf{S}}={\bf I}+\bm{\mathcal{E}}. \label{S}
\ee
Using Eq. (\ref{polar}) the stress $\bm{\mathcal{E}}$ can be re-expressed through ${\bf M}$ as
\be
\bm{\mathcal{E}}=\left({\bf M}^{\text T}{\bf M}\right)^\frac{1}{2}-{\bf I}, \label{nominal}
\ee
where superscript ${\text T}$  denotes the matrix transpose. The strain $\bm{\mathcal{E}}$
is sometimes referred to as the Biot strain \cite{Basar}. 


A measure of strain most often used in  the literature on elastic constants is the Lagrangian strain $\bm{\mathcal{H}}$ given by
\be
\bm{\mathcal{H}}=\frac{1}{2}\left({\bf M}^{\text T}{\bf M}-{\bf I}\right)=\frac{1}{2}\left({\bf S}^2-{\bf I}\right)=\bm{\mathcal{E}}+\frac{\bm{\mathcal{E}}^2}{2}, \label{Lagrangian}
\ee
for which 
\be
{\bf S}=\left({\bf I}+2\bm{\mathcal{H}}\right)^{\frac{1}{2}}. \label{SH}
\ee

In this work we choose to define elastic constants using $\bm{\mathcal{E}}$ rather than $\bm{\mathcal{H}}$ for the following reasons.
First, it is more natural to
study system response to linear changes in geometry
as given by Eq. (\ref{S}) rather than the nonlinear ones as given by Eq. (\ref{SH}). Second, because of the linearity of Eq. (\ref{S}) the use of $\bm{\mathcal{E}}$ leads to simpler general expressions for the elastic
constants. Third, there exists a one-to-one correspondence
between the elastic constants defined using $\bm{\mathcal{E}}$
and those defined using $\bm{\mathcal{H}}$ (see Appendix \ref{Ap1}). Thus one set of constants can  always be converted to the other one if necessary. (Note that for pair potentials the use of $\bm{\mathcal{H}}$ rather than $\bm{\mathcal{E}}$ leads to slightly simpler microscopic expressions for elastic constants. This is generally not true for more complex potentials.)

\subsection{Isothermal elastic constants}

The second-order isothermal elastic constants can be defined using Helmholtz free energy $A$ as \cite{Wallace}
\be
C^T_{\alpha\beta\mu\nu}=\frac{1}{V}\left(\frac{\partial^2 A}{\partial \varepsilon_{\alpha\beta}\partial\varepsilon_{\mu\nu}}\right)_{T,\bm{\mathcal{E}}=0}. \label{CT}
\ee
Here $V$ is the system volume and 
 $\varepsilon_{\alpha\beta}$ denotes the component of the symmetric strain tensor $\bm{\mathcal{E}}$; subscripts $T$ and $\bm{\mathcal{E}}=0$ mean that the derivatives are evaluated at constant temperature and for $\bm{\mathcal{E}}=0$.
Since the tensor $\bm{\mathcal{E}}$ is symmetric, the derivative with respect to $\varepsilon_{\alpha\beta}$ in Eq. (\ref{CT})
means $(\partial/\partial\varepsilon_{\alpha\beta}+\partial/\partial\varepsilon_{\beta\alpha})/2$. 

The Helmholtz free energy in Eq. (\ref{CT}) is given by
\be
A=-k_BT\ln Z, \label{Helmholtz}
\ee
where
\be
Z=\int d{ {\bf p}}^{3N}d{{\bf x}}^{3N} e^{-\frac{H}{k_BT}}.\label{Z} 
\ee
Here the integration extends over the whole phase space.
The dependence of $A$ on strain comes from the dependence of the integral in 
Eq. (\ref{Z}) on integration limits in coordinate space. A standard 
technique to avoid the mathematically inconvenient strain-dependence of the integration limits is to transfer this dependence to the Hamiltonian explicitly while holding the integration limits fixed. This can be achieved by using a suitable variable transformation.
This approach was first used by Bogoliubov 
to derive the microscopic expression for pressure as
a derivative of $A$ with respect to volume  \cite{Bogoliubov}. It
was later generalized to express the dependence of $A$ on the full strain tensor \cite{Hoover1,Parrinello82,Ray88}.

Here we use the following strain-dependent canonical transformation of coordinates and momenta of the Hamiltonian (\ref{Hamiltonian}) 
\be
\widetilde {\bf x}_{{\bf n}i}=\left({\bf I}+\bm{\mathcal{E}}\right){\bf x}_{{\bf n}i}, \qquad \widetilde {\bf p}_i=\left({\bf I}+\bm{\mathcal{E}}\right)^{-1}{\bf p}_i. \label{canon}
\ee
Note that the transformation of coordinates affects both the atoms in the central box and the image boxes.
 Inserting $\widetilde {\bf p}_i$, $\widetilde {\bf x}_{{\bf n} i}$ in the Hamiltonian (\ref{Hamiltonian}) leads to the strain-dependent transformed Hamiltonian $\widetilde H=\widetilde H_0+\widetilde U$, and the strain-dependent
 Helmholtz free energy $A$  for which the derivatives with respect to the strain components can now be evaluated.
 Using Eqs. (\ref{CT}, \ref{Helmholtz}, \ref{Z}) we obtains
\be
C_{\alpha\beta\mu\nu}^T=F_{\alpha\beta\mu\nu}^T+K_{\alpha\beta\mu\nu}^T+B_{\alpha\beta\mu\nu}^T, \label{CTT}
\ee
where
\be
F_{\alpha\beta\mu\nu}^T=\frac{1}{k_BTV}\left(\left\bra\frac{\partial\widetilde{ H}}{\partial \varepsilon_{\alpha\beta}}\right\ket\left\bra\frac{\partial \widetilde{ H}}{\partial \varepsilon_{\mu\nu}}\right\ket-\left\bra\frac{\partial \widetilde{ H}}{\partial \varepsilon_{\alpha\beta}}\frac{\partial \widetilde{ H}}{\partial \varepsilon_{\mu\nu}}\right\ket\right) \label{fluc}
\ee
is the so-called fluctuation term, 
\be
K_{\alpha\beta\mu\nu}^T=\frac{1}{V}\left\bra\frac{\partial^2 \widetilde H_0}
{\partial \varepsilon_{\alpha\beta}\partial \varepsilon_{\mu\nu}} \right\ket \label{kin}
\ee
is the kinetic term, and 
\be
B_{\alpha\beta\mu\nu}^T=\frac{1}{V}\left\bra\frac{\partial^2 \widetilde U}
{\partial \varepsilon_{\alpha\beta}\partial \varepsilon_{\mu\nu}}  \right\ket \label{B} 
\ee
is the Born term.
The derivatives in Eqs. (\ref{fluc}, \ref{kin}, \ref{B}) are evaluated at constant $T$ and for $\bm{\mathcal{E}}=0$ as in Eq. (\ref{CT}) with the corresponding subscripts dropped for brevity; the brackets denote averaging over the canonical ensemble.

The derivatives of $\widetilde H_0$ and $\widetilde U$ with respect to strain components in Eqs. (\ref{fluc}, \ref{kin}, \ref{B}) can be rewritten in terms of derivatives with respect to momenta and coordinates using Eq. (\ref{canon}).
 Calculation of these derivatives is rather straightforward 
 for $\widetilde U$ because of the linear dependence of $\widetilde {\bf x}_{{\bf n} i}$
 on $\bm{\mathcal{E}}$. The derivatives of $\widetilde H_0$ can also be calculated easily
 if one recalls that $\sum_{\alpha}\widetilde p_{i\alpha}^2$ is a dot product of vector $\widetilde {\bf p}_i$ with itself and 
 \bea
 \widetilde {\bf p}_i\cdot\widetilde {\bf p}_i&=&{\bf p}_i\cdot\left[\left(({\bf I}+\bm{\mathcal{E}})^{-1}\right)^2{\bf p}_i\right] \non \\
 &=&{\bf p}_i\cdot\left[({\bf I}-2\bm{\mathcal{E}}+3\bm{\mathcal{E}}^{2}){\bf p}_i\right]+O(\bm{\mathcal{E}}^3). \label{psq}
 \eea
 
Calculating the derivatives explicitly in Eq. (\ref{fluc}) we obtain for the fluctuation term
\be
F_{\alpha\beta\mu\nu}^T=-\frac{V}{k_BT}\big(\left\bra\sigma_{\alpha\beta}\sigma_{\mu\nu}\right\ket-\left\bra\sigma_{\alpha\beta}\right\ket\left\bra\sigma_{\mu\nu}\right\ket\big), \label{fluc2}
\ee
where
\bea
&&\sigma_{\alpha\beta}=
-\frac{1}{V}\sum_{i=1}^N\frac{p_{i\alpha}p_{i\beta}}{m_i} \non \\
&&+\frac{1}{2V}\sum_{{\bf n}}\sum_{i=1}^N\left( x_{{\bf n}i\alpha}\frac{\partial  U}{\partial  x_{{\bf n}i\beta}}+x_{{\bf n}i\beta}\frac{\partial  U}{\partial  x_{{\bf n}i\alpha}}\right) \label{sigma}
\eea
is the microscopic stress tensor. 
The second line of the last equation is the symmetrized form of the virial tensor (divided by volume) for a system with periodic
boundaries which was derived by Thompson et al. \cite{Thompson} without using strain derivatives explicitly. This form of the virial was referred to as the atom form in Ref.  \cite{Thompson}.

Using Eqs. (\ref{kin},\ref{psq}) the kinetic term is calculated to be
\be
K_{\alpha\beta\mu\nu}^T=\frac{3Nk_BT}{2V}
(\delta_{\alpha\mu}\delta_{\beta\nu}+\delta_{\alpha\nu}\delta_{\beta\mu}).\label{kin2}
\ee
Note that the kinetic term (\ref{kin2}) is $\frac{3}{4}$ of the kinetic term defined using the Lagrangian strain \cite{Gao}. We will use this fact in subsection \ref{silicon} when comparing the elastic constants defined using the Biot and Lagrangian strains.

The Born term is 
\bea
&&B_{\alpha\beta\mu\nu}^T=\non \\&&\frac{1}{V}\sum_{{\bf n},{\bf m}}\sum_{i,j=1}^N\left\bra x_{{\bf n}i\alpha} x_{{\bf m}j\mu}\frac{\partial^2  U}{\partial  x_{{\bf n}i\beta}\partial  x_{{\bf m}j\nu}}\right\ket_{{\text{sym}}}. \label{Born1}
\eea
Here the subscript sym is introduced to shorten the corresponding expression. It means that the expression in brackets is the average of four terms, namely the term shown and the three terms obtained from it by the following subscript changes $(\alpha\leftrightarrow\beta)$, $(\mu\leftrightarrow\nu)$, and $(\alpha\leftrightarrow\beta,\mu\leftrightarrow\nu)$.
The last equation represents the atom form of the Born term for a system with periodic boundaries.

  For practical 
 applications it is useful to derive the equivalent atomic-group forms of 
 Eqs. (\ref{sigma},\ref{Born1}). These can be obtained from the atomic-group 
 form of the Hamiltonian (\ref{Ham2}) by using the canonical transformation
 \be
\widetilde {\bf x}_{i}^k=\left({\bf I}+\bm{\mathcal{E}}\right){\bf x}_{i}^k, \qquad \widetilde {\bf p}_i=\left({\bf I}+\bm{\mathcal{E}}\right)^{-1}{\bf p}_i. \label{canon2}
\ee
and repeating the steps that lead to Eqs. (\ref{sigma},\ref{Born1}). This
gives
\be
\sigma_{\alpha\beta}=
-\frac{1}{V}\sum_{i=1}^N\frac{p_{i\alpha}p_{i\beta}}{m_i} 
-\frac{1}{V}\sum_{k}W_{\alpha\beta}^k, \label{sigma2}
\ee
where 
\be
W_{\alpha\beta}^k=-\frac{1}{2}\sum_{i=1}^{N_k}\left( x_{i\alpha}^k\frac{\partial  u_k}{\partial  x_{i\beta}^k}+x_{i\beta}^k\frac{\partial  u_k}{\partial  x_{i\alpha}^k}\right) \label{vir1}
\ee
is the symmetrized form of the atomic-group virial tensor for a system with periodic
boundaries which was also derived in Ref. \cite{Thompson} without
using strain derivatives explicitly. Thus, the atomic-group virial $W_{\alpha\beta}^k$ given by Eq. (\ref{vir1}) is expressed  
 in terms of  
 coordinate components of atoms in a given group and the partial force components due to the group potential energy $u_k$.

For the atomic-group form of the Born term we obtain 
\be
B_{\alpha\beta\mu\nu}^T=
\frac{1}{V}\sum_{k}\sum_{i,j=1}^{N_k}\left\bra x_{i\alpha}^k x_{j\mu}^k\frac{\partial^2  u_k}{\partial  x_{i\beta}^k\partial  x_{j\nu}^k}\right\ket_{\text{sym}}.\label{Born2}
\ee
 Equation (\ref{Born2}) can be viewed as an extension of Eq. (\ref{vir1})  for the group-form of the virial for periodic systems: the virial is obtained from the first derivatives of $\widetilde U$ with respect to strain components whereas Eq. (\ref{Born2}) is derived using the second derivatives. Equation (\ref{Born2}) can be used directly to calculate
the Born term contribution to the elastic constants: 
once the potential energy in the group form is 
expressed in Cartesian coordinates the second derivatives in 
Eq. (\ref{Born2}) can be calculated. These expressions 
can be complicated for potentials that involve few-particle interactions.

Here we show that the  right-hand
side of Eq. (\ref{Born2}) can be reduced
to a form that involves only first derivatives by using integration by parts with respect to one of the derivatives.
The basic steps of the derivation are similar to those used in the derivation
of the
equipartition theorem \cite{Tolman,Pathria}. A similar approach was used in Ref. \cite{PereverzevSewell2015} to express an ensemble-averaged matrix of second derivatives in terms of the force-force covariance matrix.  
The details of the derivation are given in Appendix \ref{Ap2}. The final expression, which is the central result of this work, is 
\bea
B_{\alpha\beta\mu\nu}^T
&=&\frac{1}{2k_BTV}\sum_{k}\left\bra \left(\widehat W^k_{\alpha\beta} W^k_{\mu\nu}+
W^k_{\alpha\beta} \widehat W^k_{\mu\nu}\right)\right\ket\non \\
&+&\frac{1}{2V}\sum_k (N_k-1)\left(\delta_{\alpha\beta}\left\bra W_{\mu\nu}^k\right\ket+\delta_{\mu\nu}\left\bra W_{\alpha\beta}^k\right\ket\right) \non \\
&+&\frac{1}{V}\sum_{k}\left(\delta_{\alpha\mu}\left\bra W^k_{\beta\nu}\right\ket\right)_{\text{sym}}, \label{Born3}
\eea
 where the  virial  $ \widehat W_{\alpha\beta}^k$ is given by
 \bea
 \widehat W_{\alpha\beta}^k&=&-\frac{1}{2}\sum_{i=1}^{N_k}\left(x_{i\alpha}^k\frac{\partial U}{\partial x_{i\beta}^k}+x_{i\beta}^k\frac{\partial U}{\partial x_{i\alpha}^k}\right)\non \\
 &&-\frac{1}{2}
 \left(\overline x_{\alpha}^k\overline f_{\beta}^k+\overline x_{\beta}^k\overline f_{\alpha}^k\right),  \label{vir2}
 \eea
 with
 \be
 \overline x_{\alpha}^k=\frac{1}{N_k}\sum_{i=1}^{N_k} x_{i \alpha}^k,\qquad 
 \overline f_{\alpha}^k=-\sum_{i=1}^{N_k} \frac{\partial U}{\partial x_{i\alpha}^k}. \label{extra}
 \ee
  Comparison of $\widehat W_{\alpha\beta}^k$ given by Eq. (\ref{vir2}) and $W_{\alpha\beta}^k$ given by Eq. (\ref{vir1}) shows
 that $\widehat W_{\alpha\beta}^k$ depends on the same 
 atomic coordinate components but the partial forces are replaced with 
 the total force component for a given atom.
 The second line of Eq. (\ref{vir2}) involves products of the arithmetic average of coordinate components of all atoms in the group
 and the sum of force components for atoms in the group; these terms ensure that $\widehat W_{\alpha\beta}^k$ is translationally invariant. 
 Translational invariance of $\widehat W_{\alpha\beta}^k$ in Eq. (\ref{vir2}) can be verified by displacing the system along any of the Cartesian directions and
 using  Eqs. (\ref{extra}) together with the translational invariance of the total force components.
 Thus, the Born term given by Eq. (\ref{Born3}) depends on the first derivatives of either atomic-group or total potential energies but does not contain the second derivatives. Because of this fact, Eq. (\ref{Born3}) can be of interest for
 numerical calculation of elastic constants, particularly for complex potentials
 that involve few-particles interaction terms.

 \subsection{Adiabatic elastic constants}
 Adiabatic elastic constants are defined using system internal energy 
 $E$ as \cite{Wallace}
 \be
C^S_{\alpha\beta\mu\nu}=\frac{1}{V}\left(\frac{\partial^2 E}{\partial \varepsilon_{\alpha\beta}\partial\varepsilon_{\mu\nu}}\right)_{S,\bm{\mathcal{E}}=0}, \label{CS}
 \ee
 where the derivatives are now evaluated at constant entropy $S$
 and for $\bm{\mathcal{E}}=0$.
 It was shown in Refs. {\cite{Graben,RR84,Ray88}} that the adiabatic elastic constants $C_{\alpha\beta\mu\nu}^S$ can be 
 calculated using the same expressions as for the isothermal 
 ones but with all canonical-ensemble averages replaced with
 averages over the microcanonical ensemble.
 Thus, Eqs. (\ref{CTT},\ref{fluc},\ref{kin},\ref{B},\ref{fluc2},\ref{kin2},\ref{Born1},\ref{Born2}) remain valid for adiabatic constants with the superscript $S$ replacing
 the superscript $T$ 
 and the 
brackets now denoting microcanonical averaging.
We show in Appendix \ref{Ap3} that Eq. (\ref{Born3}) also remains valid for the adiabatic case (with averages performed over the microcanonical ensemble). Thus, adiabatic elastic
constants can also be calculated using atomic-group virials.
 
 \section{Numerical verification}
 
  The main goal of our numerical verification is to confirm that elastic constants calculated using Eq. (\ref{CTT}) with the original Born term given by Eq. (\ref{Born2}) and the Born term calculated using virials
 (\ref{Born3}) are indeed numerically identical. To this end, we calculated elastic constants for crystalline argon and silicon 
 using molecular dynamics (MD). All MD simulations were performed using the LAMMPS package \cite{Plimpton}.
 \subsection{Isothermal elastic constants of argon}
 In the case of argon isothermal elastic constants were calculated for several values of pressure and temperature.
 The results for adiabatic constants for argon are qualitatively similar and will not be reported here. Solid argon forms a face-centered cubic (fcc) lattice.
 There are three independent elastic constants for cubic crystals \cite{Ashcroft},
 which are commonly chosen to be $C_{1111}^T, C_{1122}^T$, and $C_{1212}^T$ whose values we report below.

\begin{figure}
 \includegraphics[width=\columnwidth]{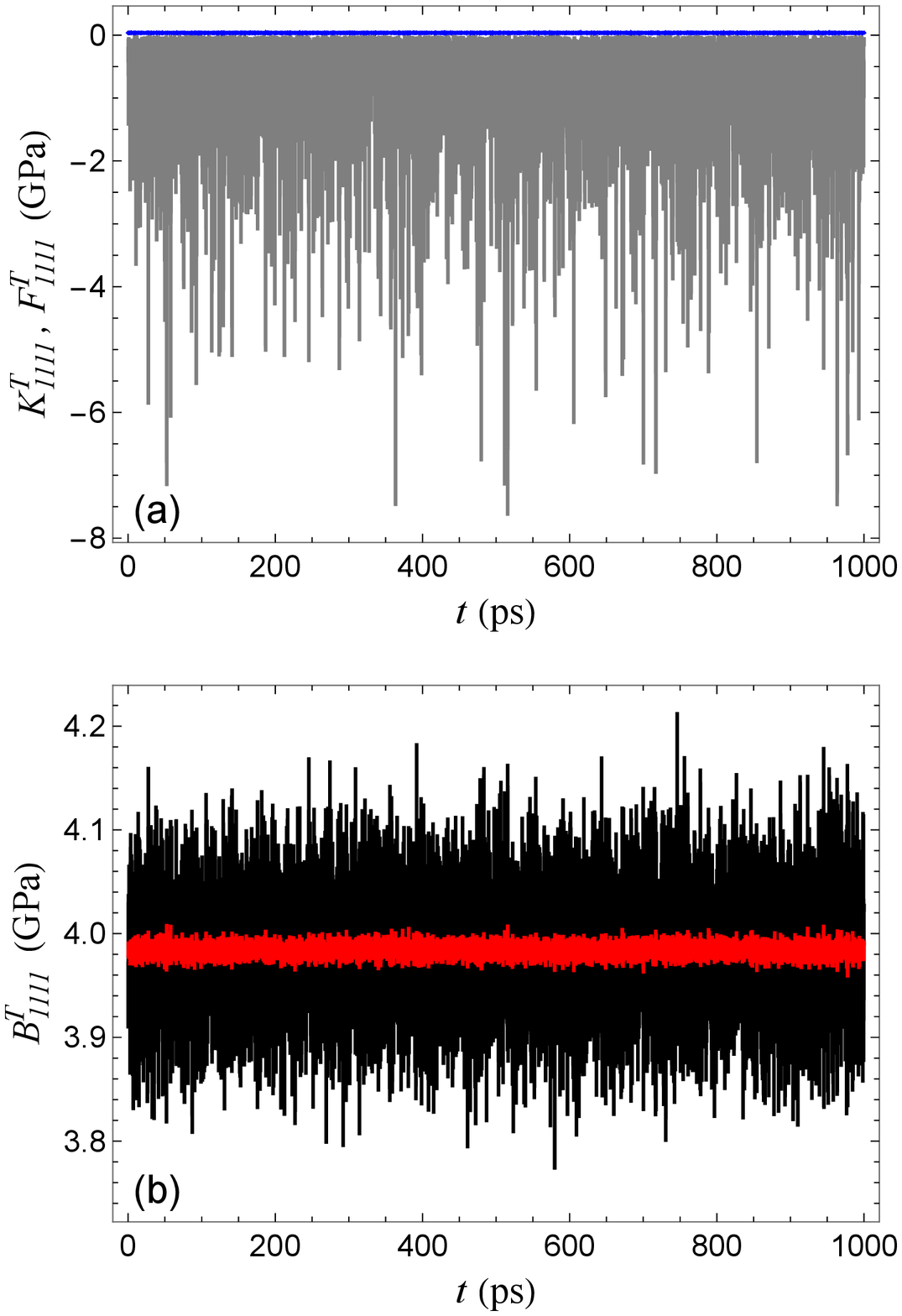}  
 \caption{\label{Figure1} Constitutive components of
$C_{1111}^T$  at zero pressure and $33\,\,{\text{K}}$ as functions of time: (a) $K_{1111}^T$ (blue) and 
$F_{1111}^T$ (gray),  (b) $B_{1111}^T$ calculated using  
Eqs. (\ref{Born2}) (red) and (\ref{Born3})  (black). Note different scales for the $y$-axis for the two panels.}
 \end{figure}

 The atomic interactions in argon were modeled with the
 Lennard-Jones potential 
\be
V(r)=4\epsilon\left[\left(\frac{\sigma}{r}\right)^{12}-\left(\frac{\sigma}{r}\right)^{6}\right] \label{LJ},
\ee
where $r$ is the distance between the two atoms, the parameters $\sigma$ and $\epsilon$ were set to be
$3.4$ {\AA} and $1.67 \times 10^{-21} $ J, respectively \cite{Bernardes,Ashcroft}. The cubic simulation cell consisting of $12\times12\times12$ conventional four-atom fcc unit cells was used. The cutoff value for the potential was 
set at the relatively high value of 28 {\AA} to minimize any possible cutoff effects.
A time step of 0.5 fs was used. Atomic coordinates were
recorded every 100 fs. These coordinates were used in Eqs.
(\ref{Born2}) and (\ref{Born3}) to calculate two forms of the Born term as functions of time. The stress tensor 
components for the fluctuation term were calculated using 
LAMMPS.
 Elastic constants were calculated at zero pressure and at $P=1$ GPa. For each of the two pressures three temperature values were considered.  These temperature  values corresponded, approximately, to
 2\%, 40\%, and 90\% of the argon melting temperature at a given pressure, which we will refer to as the low, moderate and high temperature, respectively. The melting temperature of argon is approximately 84 K at zero pressure \cite{crc} and 250 K at 1 GPa \cite{Stishov}. Thus, the temperatures of 2 K, 33 K, and 75 K for zero pressure and  5 K, 100 K, and 225 K 
 for the 1 GPa pressure were considered. The fluctuation terms, the kinetic term, and the two forms of the Born term were  
 calculated from 1 ns equilibrium NVT trajectory for each set of pressure and temperature values.
 As an example,  the behavior of these terms  as functions of time for $C_{1111}^T$ is 
 shown in Fig. \ref{Figure1} for the case of zero pressure and $T=33$ K.
 
 Tables \ref{tab} and \ref{tab2} list
 the elastic constants 
 along with the kinetic, fluctuation, and Born terms for each set of pressure and temperature values. In the case of the Born terms two values are listed: the first line shows the values obtained from Eq. (\ref{Born2}) and the second line gives the values obtained from Eq. (\ref{Born3}). The two values for the elastic
 constants in Tables \ref{tab} and \ref{tab2} are calculated using the two forms of the Born term
 with the same kinetic and fluctuation terms. The uncertainties  
 in Tables \ref{tab} and \ref{tab2} correspond to
 one-sigma standard error. They were calculated 
  using statistical analysis
 for the time-dependent correlated data \cite{Flyvbjerg}. 
 
 \begin{table*}[h!]
\caption{Isothermal elastic constants of  argon at zero pressure. The constitutive kinetic, fluctuation, and Born terms are also shown. The two values for the Born term are obtained from  Eqs. 
(\ref{Born2}) and (\ref{Born3}). The units are GPa. \label{tab}}
\begin{ruledtabular}
\begin{tabular}{cccc}
  & 2 K & 33 K & 75 K \\
\hline
\rule{0pt}{3ex}
\multirow{1}{4em}{$K_{1111}^T$}& $0.0022947\pm0.0000003 $& $0.036483\pm0.000005$ & $0.07742\pm0.00001$\\
\rule{0pt}{4ex}
\multirow{1}{4em}{$F_{1111}^T$}& $-0.0440\pm0.0004$ & $-0.695\pm0.006$ & $-1.50\pm0.01$ \\
\rule{0pt}{4ex}
\multirow{2}{4em}{$B_{1111}^T$}& $4.4324\pm0.0001$ & $3.9840\pm0.0003$ &$ 3.3035\pm0.0003 $\\
& $4.43\pm0.01$ & $3.981\pm0.003$ & $3.302\pm0.001$\\
\rule{0pt}{4ex}
\multirow{2}{4em}{$C_{1111}^T$}&$4.3907\pm0.0004$& $3.326\pm0.006$ & $1.88\pm0.01$\\
&$4.38\pm0.01$  & $3.323\pm0.007 $& $1.88\pm0.01$\\
\hline
\rule{0pt}{3ex}
\multirow{1}{4em}{$K_{1122}^T$}& $0 $& $0$ & $0$\\
\rule{0pt}{4ex}
\multirow{1}{4em}{$F_{1122}^T$}& $-0.0268\pm0.0003$ & $-0.381\pm0.005$ & $-0.80\pm0.01$ \\
\rule{0pt}{4ex}
\multirow{2}{4em}{$B_{1122}^T$}& $2.5339\pm0.0003$ & $2.2525\pm0.0001$ &$ 1.8146\pm0.0001 $\\
& $2.53\pm0.03$ & $2.252\pm0.001$ & $1.814\pm0.001$\\
\rule{0pt}{4ex}
\multirow{2}{4em}{$C_{1122}^T$}&$2.5071\pm0.0004$ & $1.872\pm0.005$ & $1.02\pm0.01$\\
& $2.50\pm0.03$ & $1.871\pm0.005$ & $1.02\pm0.01$\\
\hline
\rule{0pt}{3ex}
\multirow{1}{4em}{$K_{1212}^T$}& $0.0011473\pm0.0000002 $& $0.018242\pm0.000002$ & $0.038708\pm0.000005$\\
\rule{0pt}{4ex}
\multirow{1}{4em}{$F_{1212}^T$}& $-0.0184\pm0.0002$ & $-0.306\pm0.002$ & $-0.625\pm0.005$ \\
\rule{0pt}{4ex}
\multirow{2}{4em}{$B_{1212}^T$}& $2.5343\pm0.0003$ & $2.2586\pm0.0001$ &$ 1.8276\pm0.0001 $\\
& $2.53\pm0.04$ & $2.258\pm0.001$ & $1.827\pm0.001$\\
\rule{0pt}{4ex}
\multirow{2}{4em}{$C_{1212}^T$}&$2.5170\pm0.0004$ & $1.971\pm0.002$ &$1.241\pm0.005$\\
&$2.51\pm0.04$& $1.970\pm0.002$ & $1.241\pm0.005$\\
\end{tabular}
\end{ruledtabular}
\end{table*}

\begin{table*}[h!]
\caption{Isothermal elastic constants of argon at $P=1$ GPa. The constitutive kinetic, fluctuation, and Born terms are also shown. The two values for the Born term are obtained from  Eqs. 
(\ref{Born2}) and (\ref{Born3}). The units are GPa. \label{tab2} }
\begin{ruledtabular}
\begin{tabular}{cccc}
  & 5 K & 100 K & 225 K \\
\hline
\rule{0pt}{3ex}
\multirow{1}{4em}{$K_{1111}^T$}& $0.006753\pm0.000001$& $0.13024\pm0.00001$ & $0.27713\pm0.00004$\\
\rule{0pt}{4ex}
\multirow{1}{4em}{$F_{1111}^T$}& $-0.112\pm0.001$ & $-2.08\pm0.02$ & $-4.43\pm0.04$ \\
\rule{0pt}{4ex}  
\multirow{2}{4em}{$B_{1111}^T$}& $13.2845\pm0.0001$ & $12.5129\pm0.0004$ &$ 11.5724\pm0.0005 $\\
& $13.29\pm0.02$ & $12.512\pm0.002$ & $11.572\pm0.001$\\
\rule{0pt}{4ex}
\multirow{2}{4em}{$C_{1111}^T$}&$13.180\pm0.001$& $10.56\pm0.02$ & $7.42\pm0.04$\\
&$13.18\pm 0.02 $& $10.56\pm0.02$& $7.42\pm0.04$\\
\hline
\rule{0pt}{3ex}
\multirow{1}{4em}{$K_{1122}^T$}& $0 $& $0$ & $0$\\
\rule{0pt}{4ex}
\multirow{1}{4em}{$F_{1122}^T$}& $-0.0618\pm0.0007$ & $-1.09\pm0.01$ & $-2.00\pm0.03$ \\
\rule{0pt}{4ex}
\multirow{2}{4em}{$B_{1122}^T$}& $7.6040\pm0.0001$ & $7.0648\pm0.0003$ &$ 6.3332\pm0.0005 $\\
& $7.605\pm0.008$ & $7.065\pm0.001$ & $6.333\pm0.002$\\
\rule{0pt}{4ex}
\multirow{2}{4em}{$C_{1122}^T$}&$7.5422\pm0.0007$ &$5.98\pm0.01$& $4.33\pm0.03$\\
&$7.543\pm0.008$ & $5.98\pm0.01$& $4.33\pm0.03$\\
\hline
\rule{0pt}{3ex}
\multirow{1}{4em}{$K_{1212}^T$}& $0.0033765\pm0.0000005 $& $0.065118\pm0.000006$ & $0.13856\pm0.00002$\\
\rule{0pt}{4ex}
\multirow{1}{4em}{$F_{1212}^T$}& $-0.0489\pm0.0004$ & $-0.898\pm0.007$ & $-1.85\pm0.02$ \\
\rule{0pt}{4ex}
\multirow{2}{4em}{$B_{1212}^T$}& $7.1051\pm0.0001$ & $6.5865\pm0.0002$ &$5.8795\pm0.0005$\\
& $7.11\pm0.01$ & $6.587\pm0.001$ & $5.880\pm0.002$\\
\rule{0pt}{4ex}
\multirow{2}{4em}{$C_{1212}^T$}&$7.0595\pm0.0004$& $5.753\pm0.007$ & $4.16\pm0.02$\\
& $7.06\pm0.01$ & $5.754\pm0.007$ & $4.16\pm0.02$\\
\end{tabular}
\end{ruledtabular}
\end{table*}

One can see from Tables \ref{tab} and \ref{tab2} that
elastic constants calculated using two 
forms of the Born term are indeed the same within uncertainties for all pressure and temperature values considered. 
Note that in all cases the Born term calculated  using Eq. (\ref{Born3}) converges 
slower than the one obtained from the original expression (\ref{Born2}), as reflected by 
the corresponding uncertainties in Tables \ref{tab} and \ref{tab2}. (See also  panel (b) of Fig. \ref{Figure1}). However, for moderate and high temperatures the Born term 
in the form of Eq. (\ref{Born3}) converges faster than the fluctuation term (cf. Fig. \ref{Figure1}), whereas for low temperatures it converges slower than the fluctuation term.
Thus, for moderate and high temperatures the overall convergence of the elastic constants 
 calculated using the Born term  given by Eq. (\ref{Born3}) is approximately the same as that for the elastic constants 
 obtained using the Born term  given by Eq. (\ref{Born2}). For low temperatures
 the elastic constants 
 calculated using  Eq. (\ref{Born3}) converge slower.
\subsection{Adiabatic elastic constants of silicon} \label{silicon}
In the case of silicon, adiabatic rather than isothermal elastic constants were calculated. This was done for the following two reasons: to show that the new expression for the Born term can be also be applied to obtain adiabatic constants and to compare the calculated values to earlier MD-based  adiabatic results for elastic constants of silicon \cite{Kluge,ZhenChu}.
Similarly to fcc argon, silicon is a cubic crystal. Thus, only $C^S_{1111}$, $C^S_{1122}$, and $C^S_{1212}$ needed to be calculated. The system was studied at zero pressure and the following four temperatures:
300 K, 888 K, 1164 K, and 1477 K. The latter three temperatures corresponded to the ones considered in Refs. \cite{Kluge,ZhenChu}.

The atomic interactions in silicon were modeled with the Stillinger-Weber potential \cite{SW} which consists of two- and three-atom interaction terms. A cubic simulation cell of 216 silicon atoms was used. All force-field and other simulation parameters reported in Refs. \cite{Kluge,ZhenChu} were kept except for the total number of simulation steps. References   \cite{Kluge} and \cite{ZhenChu} used 150000 
and 1 million steps, respectively.  We used 2 million steps for the three higher temperatures and 6 million steps for 300 K to obtain good convergence of the Born terms obtained using Eq. (\ref{Born3}).

 Kinetic, fluctuation, and Born terms were calculated using microcanonical averaging, i. e. NVE trajectories, for each of the four temperatures. Two forms of the Born terms were calculated: the original one using Eq. (\ref{Born2}) and the new form based on Eq. (\ref{Born3}). For each of these two forms both two-atom and three-atom contributions to the total Born terms were calculated. These results are summarized in Table \ref{tab3}. The two values for the elastic constants reported in Table \ref{tab3} were calculated using the two forms of the total Born term with the same kinetic and fluctuation terms.  

One can see that
elastic constants calculated using two 
forms of the Born term agree with each other within uncertainties for all temperatures considered. 
This remains true for the two-atom, three-atom, and total Born terms calculated using the two approaches.
For all cases the Born terms calculated  using Eq. (\ref{Born3}) converges 
slower than the ones obtained from the original expression (\ref{Born2}), as reflected by 
the corresponding uncertainties. Comparison of uncertainties for the fluctuation terms and the Born terms calculated from Eq. (\ref{Born3}) shows that $B_{1111}^S$ and $B_{1122}^S$ converge slower and  $B_{1212}^S$ much faster than the corresponding fluctuation terms for all four temperatures.

The results of Ref. \cite{ZhenChu} for the elastic constants of silicon along with the constitutive Born, kinetic, and fluctuation terms for 888 K, 1164 K, and 1477 K are also listed in Table \ref{tab3}. Also shown are the 
elastic constants reported in Ref. \cite{Kluge} for the same three temperatures.
When comparing our results to the previously  published data we need to bear in mind that  the elastic constants reported in Refs. \cite{Kluge,ZhenChu} were calculated using the Lagrangian strain whereas our results were based of the Biot strain. Fortunately, the transformation  between the Lagrangian-strain and Biot-strain elastic constants (and their constitutive terms) is straightforward. It follows from Eq. (\ref{rela}) that at zero pressure the adiabatic elastic constants 
obtained using the Lagrangian and Biot strains are identical, thus they can be compared directly. However, this is not true for the kinetic and Born terms for the two strains considered  separately. It was noted following Eq. (\ref{kin2}) that the Biot kinetic term is $\frac{3}{4}$ of the Lagrangian one. The fluctuation terms for the two strain measures are identical because the stress tensors obtained using the two strains are identical. This implies that the Biot Born term is equal the Lagrangian Born term plus one fourth of the Lagrangian kinetic term. Thus, the original Lagrangian results of Ref. \cite{ZhenChu} for the kinetic and total Born terms were converted to the Biot form using the recipe just above when reporting them in Table \ref{tab3}. 

One can see that the total Born terms of Ref. \cite{ZhenChu} and both forms of the total Born terms 
calculated by us are generally in  good agreement: for most cases they are identical within uncertainties. Elastic constants calculated by us  also compare well to the elastic constants  from Refs. \cite{Kluge} and \cite{ZhenChu}. The larger differences for $C_{1212}^S$ at 1164 K and 1477 K are still within uncertainties and are due to the very large uncertainties of the corresponding fluctuation terms.

 \begin{table*}[h!]
\caption{Adiabatic elastic constants of Si at zero pressure. The constitutive kinetic, fluctuation, and two-atom, three-atom and total Born terms are also shown. The two listed values for the Born terms are obtained from  Eqs. (\ref{Born2}) and (\ref{Born3}). Results of Refs. \cite{ZhenChu} and \cite{Kluge} are also shown. The units are $10^{10}$ Pa.\label{tab3}}
\begin{ruledtabular}
\begin{tabular}{ccccc}
  & 300 K & 888 K & 1164 K & 1477 K \\
\hline
\rule{0pt}{3ex}
\multirow{1}{4em}{$K_{1111}^S$}& $0.0619\pm0.0006$& $0.1819\pm0.0005$ & $0.2378\pm0.0007$&$0.3015\pm0.0008$\\
\rule{0pt}{3ex}
\multirow{1}{4em}{$K_{1111}^S{\text{ Ref.\cite{ZhenChu}}}$}& & $0.18$ & $0.25$&$0.30$\\
\rule{0pt}{3ex}
\multirow{1}{4em}{$F_{1111}^S$}& $-0.087\pm0.002$ & $-0.345\pm0.009$ & $-0.50\pm0.01$ &$-0.70\pm0.01$\\
\rule{0pt}{3ex}
\multirow{1}{4em}{$F_{1111}^S{\text{ Ref.\cite{ZhenChu}}}$}&  & $-0.34$ & $-0.53$ &$-0.71$\\
\rule{0pt}{3ex}
\multirow{2}{4em}{$B_{1111}^S{\text{(2-atom)}}$}&$10.1571\pm0.0004 $& $9.687\pm0.006$  &$ 9.17\pm 0.01$&$8.44\pm0.02$\\
& $10.14\pm0.05$ & $9.68\pm0.06$  & $9.21 \pm 0.05$& $8.49\pm0.06$\\
\rule{0pt}{3ex}
\multirow{2}{4em}{$B_{1111}^S{\text{(3-atom)}}$}&$4.781\pm0.001$ & $4.669\pm0.004$ &$4.883\pm0.006 $&$5.31\pm0.01$\\
&$4.76\pm0.04$ & $4.68\pm0.02$ & $4.88\pm0.01$&$5.31\pm0.02$\\
\rule{0pt}{3ex}
\multirow{2}{4em}{$B_{1111}^S{\text{(total)}}$}& $14.938\pm0.001$ & $14.356\pm0.007$ &$ 14.05\pm0.01 $&$13.76\pm0.02$\\
&$14.90\pm0.06$  & $14.36\pm0.06$ & $14.09\pm0.05$& $13.81\pm0.06$\\
\rule{0pt}{3ex}
\multirow{1}{4em}{$B_{1111}^S {\text{(total) Ref.\cite{ZhenChu}}}$} &  & $14.35$ & $14.01$ & $13.76$\\
\rule{0pt}{3ex}
\multirow{2}{4em}{$C_{1111}^S$}&$14.912\pm0.002$ & $14.19\pm0.01$ & $13.79\pm0.01$&$13.35\pm0.03$\\
&$14.88\pm 0.06$  & $14.20\pm0.09 $& $13.82\pm0.05$&$13.42\pm0.06$\\
\rule{0pt}{3ex}
\multirow{1}{4em}{$C_{1111}^S{\text{ Ref.\cite{ZhenChu}}}$}&  & $14.19\pm0.02$ & $13.73\pm0.02$&$13.35 \pm 0.02$ \\
\rule{0pt}{3ex}
\multirow{1}{4em}{$C_{1111}^S{\text{ Ref.\cite{Kluge}}}$}&  & $14.14\pm0.01$ & $13.73\pm0.03$ &$13.32 \pm 0.01$\\
\hline
\rule{0pt}{3ex}
\multirow{1}{4em}{$K_{1122}^S$}& $0 $& $0$ & $0$&$0$\\
\rule{0pt}{3ex}
\multirow{1}{4em}{$F_{1122}^S$}& $0.017\pm0.002$ & $0.006\pm0.004$ & $-0.01\pm0.01$&$-0.04\pm0.01$ \\
\rule{0pt}{3ex}
\multirow{1}{4em}{$F_{1122}^S{\text{ Ref.\cite{ZhenChu}}}$}&  & $0.01$ & $-0.02$&$-0.05$ \\
\rule{0pt}{3ex}
\multirow{2}{4em}{$B_{1122}^S{\text{(2-atom)}}$}& $9.980\pm0.002$ & $9.407\pm0.006$ &$ 9.17\pm0.01 $&$8.41\pm0.01$\\
& $9.96\pm0.05$ & $9.40\pm0.06$ & $9.21\pm0.05$&$8.46\pm0.06$\\
\rule{0pt}{3ex}
\multirow{2}{4em}{$B_{1122}^S{\text{(3-atom)}}$}&$-2.353\pm0.001$  & $-1.891\pm0.005$ &$ -1.513\pm0.007 $&$-0.94\pm0.01$\\
&$-2.34\pm0.02$  & $-1.90\pm0.01$ & $-1.51\pm0.01$&$-0.95\pm0.02$\\
\rule{0pt}{3ex}
\multirow{2}{4em}{$B_{1122}^S{\text{(total)}}$}& $7.626\pm0.002$ & $7.516\pm0.008$ &$7.47\pm0.01 $&$7.47\pm0.01$\\
& $7.62\pm0.05$ & $7.50\pm0.06$ & $7.50\pm0.05$&$7.51\pm0.06$\\
\rule{0pt}{3ex}
\multirow{1}{4em}{$B_{1122}^S {\text{(total) Ref.\cite{ZhenChu}}}$}& & $7.52$ & $7.47$ & $7.48$\\
\rule{0pt}{3ex}
\multirow{2}{4em}{$C_{1122}^S$}&$7.643\pm0.003$ & $7.522\pm0.009$ & $7.46\pm0.01$&$7.42\pm0.01$\\
& $7.64\pm0.05$ & $7.51\pm0.06$ & $7.49\pm0.05$&$7.47\pm0.06$\\
\rule{0pt}{3ex}
\multirow{1}{4em}{$C_{1122}^S{\text{ Ref.\cite{ZhenChu}}}$}&  & $7.53\pm0.01$ & $7.45\pm0.01$ &$7.43\pm0.01$\\
\rule{0pt}{3ex}
\multirow{1}{4em}{$C_{1122}^S{\text{ Ref.\cite{Kluge}}}$}&  & $7.52\pm0.00$ & $7.43\pm0.01$ &$7.39\pm0.04$\\
\hline
\rule{0pt}{3ex}
\multirow{1}{4em}{$K_{1212}^S$}& $0.0309\pm0.0003 $& $0.0909\pm0.0002$ & $0.1189\pm 0.0004$&$0.1507\pm0.0004$\\
\rule{0pt}{3ex}
\multirow{1}{4em}{$K_{1212}^S{\text{ Ref.\cite{ZhenChu}}}$}&  & $0.09$ & $0.12$&$0.15$\\
\rule{0pt}{3ex}
\multirow{1}{4em}{$F_{1212}^S$}& $-5.6\pm0.4$ & $-5.0\pm0.3$ & $-5.4\pm0.2$&$-5.6\pm0.3$ \\
\rule{0pt}{3ex}
\multirow{1}{4em}{$F_{1212}^S{\text{ Ref.\cite{ZhenChu}}}$}& & $-4.92$ & $-5.66$ & $-5.87$ \\
\rule{0pt}{3ex}
\multirow{2}{4em}{$B_{1212}^S${\text{(2-atom)}}}& $10.003\pm0.001$ & $9.483\pm0.006$ &$ 9.089\pm0.007 $&$8.56\pm0.01$\\
& $9.99\pm0.05$ & $9.48\pm0.06$ & $9.12\pm0.05$&$8.61\pm0.06$\\
\rule{0pt}{3ex}
\multirow{2}{4em}{$B_{1212}^S${\text{(3-atom)}}}& $0.8285\pm0.0002$ & $0.969\pm0.002$ &$ 1.161\pm0.004$&$1.468\pm0.007$\\
& $0.82\pm0.01$ & $0.97\pm0.01$ & $1.16\pm0.01$&$1.47\pm0.01$\\
\rule{0pt}{3ex}
\multirow{2}{4em}{$B_{1212}^S${\text{(total)}}}& $10.832\pm0.001$ & $10.451\pm0.006$ &$ 10.251\pm0.007 $&$10.03\pm0.01$\\
& $10.81\pm0.05$ & $10.45\pm0.06$ & $10.28\pm0.05$&$10.07\pm0.06$\\
\rule{0pt}{3ex}
\multirow{1}{4em}{$B_{1212}^S{\text{(total) Ref.\cite{ZhenChu}}}$}&  & $10.45$ & $10.22$&$10.03$ \\
\rule{0pt}{3ex}
\multirow{2}{4em}{$C_{1212}^S$}&$5.3\pm0.4$ & $5.6\pm0.3$ &$4.9\pm0.2$&$4.6\pm0.3$\\
&$5.3\pm0.4$& $5.6\pm0.3$ & $5.0\pm0.2$&$4.7\pm0.3$\\
\rule{0pt}{3ex}
\multirow{1}{4em}{$C_{1212}^S{\text{ Ref.\cite{ZhenChu}}}$}&  & $5.62\pm0.54$ & $4.68\pm0.21$&$4.31\pm0.36$ \\
\rule{0pt}{3ex}
\multirow{1}{4em}{$C_{1212}^S{\text{ Ref.\cite{Kluge}}}$}&  & $5.24\pm0.84$ & $4.57\pm1.14$&$4.20\pm0.83$ \\
\end{tabular}
\end{ruledtabular}
\end{table*}

\section{Conclusions}
We derived expressions for isothermal and
  adiabatic elastic constants that explicitly
  incorporate the effect of periodic boundaries.
  These expressions can be used for numerical calculations of elastic constants and 
  represent the second-derivative generalizations of expressions for the virials of
  periodic systems \cite{Thompson,Bekker}.
 
  We also showed that the Born term can be expressed
  in the form that involves only the first derivatives
  of the atomic-group and total potential energies. This fact is important from a fundamental standpoint because it means that knowledge of
  atomic coordinates and suitably chosen partial and total forces along with system volume and temperature is sufficient to define isothermal and adiabatic elastic constants.  Equation (\ref{Born3}) is also of interest
  for numerical calculation of elastic constants (as was done
  for argon and silicon in this work): molecular dynamics simulation packages such as LAMMPS  can compute atomic virial tensors given by Eq. (\ref{vir1}) and calculation of the Born term using Eq. (\ref{Born3}) can be done
  with modest code modifications. 
  
  Application of Eq. (\ref{Born3}) to the Lennard-Jones argon and the Stillinger-Weber silicon shows very good agreement
 with the original Born expression given by Eq. (\ref{Born2}) but with a slower convergence. Numerical validation of
Eq. (\ref{Born3}) for the cases of more complex potentials along
with the studies of the system size, stress, and temperature dependence is intended by us in the near future.
  
  \begin{acknowledgments}
This research was funded by Air Force Office of Scientific Research, grant numbers FA9550-19-1-0318
and FA9550-22-1-0212, and by AFOSR DURIP equipment grant FA9550-20-1-0205. 
The author is grateful to Yuriy Pereverzev and Tommy Sewell for valuable comments and suggestions.
\end{acknowledgments}
  
\appendix
\section{}\label{Ap1}
Here we derive the relationship between the elastic constants 
$C_{\alpha\beta\mu\nu}^T$ defined using the Biot strain $\bm{\mathcal{E}}$ and the elastic
constants $\widehat{C}_{\alpha\beta\mu\nu}^T$ defined using the Lagrangian strain $\bm{\mathcal{H}}$.
We treat tensor $\bm{\mathcal{H}}$ (whose
components are denoted by $\eta_{\alpha\beta}$) as a function of $\bm{\mathcal{E}}$
as given by the last equality of Eq. (\ref{Lagrangian}) and apply the chain rule to 
express derivatives of $A$ with respect to $\varepsilon_{\alpha\beta}$ in Eq. (\ref{CT})
in terms of derivatives with respect to $\eta_{\alpha\beta}$. We have (using Einstein notation and dropping subscript $T$)
\bea
&&\left(\frac{\partial^2 A}
{\partial \varepsilon_{\alpha\beta}\partial\varepsilon_{\mu\nu}}\right)_{\bm{\mathcal{E}}=0}=
\left(\frac{\partial^2 A}{\partial \eta_{\tau\upsilon}\eta_{\phi\omega}}\frac{\partial\eta_{\tau\upsilon}}{\partial \varepsilon_{\alpha\beta}}\frac{\partial\eta_{\phi\omega}}{\partial \varepsilon_{\mu\nu}}\right.\non \\
&&\left.+\frac{\partial A}{\partial \eta_{\tau\upsilon}}\frac{\partial^2 \eta_{\tau\upsilon}}{\partial \varepsilon_{\alpha\beta}\partial\varepsilon_{\mu\nu}}\right)_{\bm{\mathcal{E}}=0}. \label{A1}
\eea
Using Eq. (\ref{Lagrangian}) we can evaluate the following derivatives
\bea
&&\left(\frac{\partial\eta_{\tau\upsilon}}{\partial \varepsilon_{\alpha\beta}}\right)_{\bm{\mathcal{E}}=0}=\frac{1}{2}
(\delta_{\tau\alpha}\delta_{\upsilon\beta}+\delta_{\tau\beta}\delta_{\upsilon\alpha} ), \non \\
&&\left(\frac{\partial^2\eta_{\tau\upsilon}}{\partial \varepsilon_{\alpha\beta}\partial \varepsilon_{\mu\nu}}\right)_{\bm{\mathcal{E}}=0}=\frac{1}{8}
\left(\delta_{\tau\alpha}\delta_{\beta\mu}\delta_{\upsilon\nu}+\delta_{\tau\beta}\delta_{\alpha\mu}\delta_{\upsilon\nu} \right. \non \\
&&\left.+\delta_{\tau\alpha}\delta_{\beta\nu}\delta_{\upsilon\mu}+\delta_{\tau\beta}\delta_{\alpha\nu}\delta_{\upsilon\mu}+\delta_{\alpha\nu}\delta_{\upsilon\beta}\delta_{\tau\mu}+\delta_{\nu\beta}\delta_{\upsilon\alpha}\delta_{\tau\mu} \right.\non \\
&&\left.+\delta_{\alpha\mu}\delta_{\upsilon\beta}\delta_{\tau\nu}+\delta_{\mu\beta}\delta_{\upsilon\alpha}\delta_{\tau\nu}\right).
\eea
Inserting these into Eq. (\ref{A1}) and using the fact that  $\bm{\mathcal{H}}=0$ when $\bm{\mathcal{E}}=0$  we obtain
\bea
&&\left(\frac{\partial^2 A}
{\partial \varepsilon_{\alpha\beta}\partial\varepsilon_{\mu\nu}}\right)_{\bm{\mathcal{E}}=0}=\left(\frac{\partial^2 A}
{\partial \eta_{\alpha\beta}\partial\eta_{\mu\nu}}\right)_{\bm{\mathcal{H}}=0} \non \\
&&+\frac{1}{4}\left(\frac{\partial A}
{\partial \eta_{\alpha\nu}}\delta_{\beta\mu}+\frac{\partial A}
{\partial \eta_{\beta\nu}}\delta_{\alpha\mu}+\frac{\partial A}
{\partial \eta_{\alpha\mu}}\delta_{\beta\nu}+\frac{\partial A}
{\partial \eta_{\beta\mu}}\delta_{\alpha\nu}\right)_{\bm{\mathcal{H}}=0}.\non \\
\eea
Dividing both sides of the last equation by volume and using the fact that
\be
\frac{1}{V}\left(\frac{\partial A}{\partial \eta_{\alpha\nu}}\right)_{\bm{\mathcal{H}}=0}=\frac{1}{V}\left(\frac{\partial A}{\partial \varepsilon_{\alpha\nu}}\right)_{\bm{\mathcal{E}}=0}=\bra\sigma_{\alpha\beta} \ket
\ee
we obtain the sought-after relationship between the two sets of isothermal elastic constants,
\bea
C_{\alpha\beta\mu\nu}^T&=&\widehat C_{\alpha\beta\mu\nu}^T+\frac{1}{4}\big(
 \bra\sigma_{\alpha\nu}\ket\delta_{\beta\mu}+\bra\sigma_{\beta\nu}\ket\delta_{\alpha\mu}+\bra\sigma_{\alpha\mu}\ket\delta_{\beta\nu}\non\\
 &&+\bra\sigma_{\beta\mu}\ket\delta_{\alpha\nu}\big). \label{rela}
\eea
Applying a similar argument to the system energy $E$, one can verify that this relationship remains true for adiabatic constants as well.

\section{} \label{Ap2}
Here we outline the derivation of Eq. (\ref{Born3}) from Eq. (\ref{Born2}).
Before proceeding with the derivation let us note that the potential energy 
$u_k$ of a group of $N_k$ atoms is translationally invariant and, therefore, 
depends on at most $3(N_k-1)$ variables. We want to account for this translational invariance
when doing integration by parts. This can be achieved by doing an orthogonal transformation  from coordinates ${\bf x}_i$ to ${\bf y}_i$. 
For each group $k$ the transformation involves $N_k$ atomic-group coordinates ${\bf x}_i^k$ and transforms them to new $N_k$ coordinates ${\bf y}_i^k$. The transformation within this group of $N_k$ coordinates can be
arbitrary apart from the fact that one of the new vector variables (assumed to be the last here) is expressed through the arithmetic average of ${\bf x}_i^k$,
i. e. ${\bf y}_{N_k}^k=\frac{1}{\sqrt{N_k}}\sum_{i=1}^{N_k}{\bf x}_{i}^k$. For coordinates not
involved in the group $k$, we take
${\bf y}_i={\bf x}_i$. 
The following relationship between the old and new variables within the group is used in the analysis below,
\be
\sum_{i=1}^{N_k}x_{i\alpha}^k\frac{\partial  g(\{{\bf x}_i^k\})}{\partial  x_{i\beta}^k}=\sum_{i=1}^{N_k}y_{i\alpha}^k\frac{\partial  g(\{{\bf y}_i^k\})}{\partial  y_{i\beta}^k}. \label{b1}
\ee
Here $g(\{{\bf x}_i^k\})$ is an arbitrary function of the old variables ${\bf x}_i^k$ and the same symbol is used for the transformed 
function $g(\{{\bf y}_i^k\})$ of the new variables ${\bf y}_i^k$. 
Transforming to the new variables and using Eq. (\ref{b1}) the double sum over atoms in one group 
in the Born term (\ref{Born2}) can be written as
\bea
\sum_{i,j=1}^{N_k}&&\left\bra x_{i\alpha}^k x_{j\mu}^k\frac{\partial^2  u_k}{\partial  x_{i\beta}^k\partial  x_{j\nu}^k}\right\ket_{\text{sym}}\non \\ &&=\sum_{i,j=1}^{N_k-1}\left\bra y_{i\alpha}^k y_{j\mu}^k\frac{\partial^2  u_k}{\partial  y_{i\beta}^k\partial  y_{j\nu}^k}\right\ket_{\text{sym}},\label{Born4}
\eea
where summations in the last expression now run from $1$ to $N_k-1$ because $u_k$ is translationally 
invariant in the space of $N_k$ atomic-group coordinates and, therefore,  does not depend on ${\bf y}_{N_k}^k$.
Consider one term in the summand of the last expression with integrations written out explicitly 
and the superscript $k$  and the subscript sym dropped for brevity:
\be
\left\bra y_{i\alpha} y_{j\mu}\frac{\partial^2  u_k}{\partial  y_{i\beta}\partial  y_{j\nu}}\right\ket=\frac{1}{Q}\int d {\bf y}^{3N} 
e^{-\frac{U}{k_BT}}y_{i\alpha}  y_{j\mu}\frac{\partial^2  u_k}{\partial  y_{i\beta}\partial  y_{j\nu}}, \label{term0}
\ee
where $Q=\int d{\bf y}^{3N}e^{-\frac{U}{k_BT}}$.
Integrating by parts with respect to $y_{i\beta}$ gives
\bea
&&\left\bra y_{i\alpha} y_{j\mu}\frac{\partial^2  u_k}{\partial  y_{i\beta}\partial  y_{j\nu}}\right\ket \non \\
&=&\frac{1}{Q}\int d {\bf y}^{3N-1} 
e^{-\frac{U}{k_BT}}y_{i\alpha}  y_{j\mu}\frac{\partial  u_k}{\partial  y_{j\nu}}\bigg|_{y_{i\beta}=l_1}^{y_{i\beta}=l_2} \non \\
&&-\frac{1}{Q}\int d {\bf y}^{3N}\frac{\partial  u_k}{\partial  y_{j\nu}}\frac{\partial}{\partial y_{i\beta} }
\left(e^{-\frac{U}{k_BT}}y_{i\alpha}  y_{j\mu} \right) \non \\
&=&\frac{1}{Q}\int d {\bf y}^{3N-1} 
e^{-\frac{U}{k_BT}}y_{i\alpha}  y_{j\mu}\frac{\partial  u_k}{\partial  y_{j\nu}}\bigg|_{y_{i\beta}=l_1}^{y_{i\beta}=l_2} \non \\
&&-\frac{1}{Q}\int d {\bf y}^{3N}\frac{\partial  u_k}{\partial  y_{j\nu}}
e^{-\frac{U}{k_BT}}\bigg(-\frac{1}{k_BT}\frac{\partial  U}{\partial  y_{i\beta}}y_{i\alpha}  y_{j\mu} \non \\&&+\delta_{\alpha\beta}y_{j\mu} 
+\delta_{ij}\delta_{\beta\mu}y_{i\alpha}\bigg). \label{term}
\eea
Here $l_1$ and $l_2$ in the second and forth lines are the integration limits for $y_{i\beta}$. 

Up to this point the derivation is exact. The key step in deriving Eq. (\ref{Born3}) is the omission of the 
integrated part, i. e. the fourth line of Eq. (\ref{term}).
The integrated part will vanish exactly in the thermodynamic limit (as the integration limits tend to $\pm \infty$) for typical atomic-group potential functions found in solids.
More specifically, for potentials that decrease with inter-atomic distance (such as the Lennard-Jones or Coulombic potentials) $u_k$ and its derivatives vanish at infinity which makes the integrated part (the first line of Eq. (\ref{term})) vanish exactly. For potentials that grow with inter-atomic distance, such as the harmonic bond potential, the exponential factor in the integrated part of Eq. (\ref{term}) will tend to zero as the potential itself goes to infinity at the boundaries, which, again, makes the integrated part vanish exactly.
If the system is finite, the integrated part may not vanish exactly but its contribution will become 
progressively smaller as the system size increases. In the following derivation we assume 
that the system is large enough to be treated as being in the thermodynamic limit and omit the integrated part. 

Clearly, integration by parts in (\ref{term0}) can also be done with respect to $y_{j\nu}$ 
with the result 
\bea
&&\left\bra y_{i\alpha} y_{j\mu}\frac{\partial^2  u_k}{\partial  y_{i\beta}\partial  y_{j\nu}}\right\ket \non \\
&=&-\frac{1}{Q}\int d {\bf y}^{3N}\frac{\partial  u_k}{\partial  y_{i\beta}}
e^{-\frac{U}{k_BT}}\bigg(-\frac{1}{k_BT}\frac{\partial  U}{\partial  y_{j\nu}}y_{i\alpha}  y_{j\mu} \non \\&&+\delta_{\mu\nu}y_{i\alpha} +\delta_{ij}\delta_{\alpha\nu}y_{j\mu}\bigg). \label{term2}
\eea
To maintain proper symmetry of elastic constants the average of (\ref{term}) and (\ref{term2}) has to be taken. Substituting this average into (\ref{Born4}), performing summations, where it can be done explicitly, and restoring the superscript $k$ and the subscript sym we obtain 
\bea
& &\sum_{i,j=1}^{N_k-1}\left\bra y_{i\alpha}^k y_{j\mu}^k\frac{\partial^2  u_k}{\partial  y_{i\beta}^k\partial  y_{j\nu}^k}\right\ket_{\text{sym}} \non \\
&=&\frac{1}{2k_BT}\sum_{i,j=1}^{N_k-1}\left\bra y_{i\alpha}^k  y_{j\mu}^k \left(\frac{\partial  u_k}{\partial  y_{j\nu}^k}\frac{\partial  U}{\partial  y_{i\beta}^k} +\frac{\partial  u_k}{\partial  y_{i\beta}^k}\frac{\partial  U}{\partial  y_{j\nu}^k}\right)\right\ket_{\text{sym}}  \non \\
& &-\frac{1}{2}\sum_{i=1}^{N_k-1}\left\bra\big((N_k-1)\delta_{\alpha\beta}y_{i\mu}^k+\delta_{\beta\mu}y_{i\alpha}^k\big)
\frac{\partial  u_k}{\partial  y_{i\nu}^k}\right\ket_{\text{sym}} \non \\
& &-\frac{1}{2}\sum_{i=1}^{N_k-1}\left\bra\big((N_k-1)\delta_{\mu\nu}y_{i\alpha}^k+\delta_{\alpha\nu}y_{i\mu}^k\big)
\frac{\partial  u_k}{\partial  y_{i\beta}^k}\right\ket_{\text{sym}}. \label{beforetr}
\eea
We now transform Eq. (\ref{beforetr}) back to the original coordinates ${\bf x}_{i}^k$ using Eq. (\ref{b1}). The sums involving $u_k$ transform as follows
\be
\sum_{i=1}^{N_k-1}y_{i\alpha}^k \frac{\partial u_k}{\partial  y_{i\beta}^k}=\sum_{i=1}^{N_k}y_{i\alpha}^k \frac{\partial  u_k}{\partial  y_{i\beta}^k}=\sum_{i=1}^{N_k}x_{i\alpha}^k \frac{\partial  u_k}{\partial  x_{i\beta}^k}. \label{b7}
\ee
Here the summation in the second expression is extended to $N_k$ because $u_k$ does not depend on ${\bf y}_{N_k}^k$. The total potential energy $U$ in Eq. (\ref{beforetr}) does depend on ${\bf y}_{N_k}^k$, in general. Therefore, when transforming back to ${\bf x}_{i}^k$,
\bea
&&\sum_{i=1}^{N_k-1}y_{i\alpha}^k \frac{\partial  U}{\partial  y_{i\beta}^k}=\left(\sum_{i=1}^{N_k}y_{i\alpha}^k \frac{\partial U}{\partial  y_{i\beta}^k}\right)-y_{N_k\alpha}^k\frac{\partial U}{\partial  y_{N_k\beta}^k} \non \\
&&=\left(\sum_{i=1}^{N_k}x_{i\alpha}^k \frac{\partial U}{\partial  x_{i\beta}^k}\right)-\frac{1}{N_k}\left(\sum_{i=1}^{N_k} x_{i \alpha}^k\right)\left(\sum_{i=1}^{N_k} \frac{\partial U}{\partial x_{i\beta}^k}\right). \label{b8} \non \\ 
\eea
Here the last term involving the product of two sums is obtained
by transforming $y_{N_k\alpha}^k\partial U/{\partial  y_{N_k\beta}^k}$ to 
coordinates ${\bf x}_{i}^k$.
Substituting Eqs. (\ref{b7}) and (\ref{b8}) into Eq. (\ref{beforetr}), performing symmetrizations,  summing over all groups, and dividing by the volume 
lead to Eq. (\ref{Born3}).
\section{} \label{Ap3}
In this Appendix we show that the Born term (\ref{Born2}) calculated using the microcanonical ensemble,
\be
\rho=\frac{1}{\Omega}\delta(E-H),
\ee
where $\Omega=\int d {\bf p}^{3N}d {\bf x}^{3N}\delta(E-H)$,
can also be rewritten in the form of Eq. (\ref{Born3}), in which all averages are taken with the microcanonical ensemble as well. 
The initial steps are the same as in Appendix \ref{Ap2} but with the canonical averages replaced
with the microcanonical ones. Thus we have the following microcanonical 
analogue of (\ref{term0})
\bea
&&\left\bra y_{i\alpha} y_{j\mu}\frac{\partial^2  u_k}{\partial  y_{i\beta}\partial  y_{j\nu}}\right\ket \non \\
&&=\frac{1}{\Omega}\int d {\bf p}^{3N} d {\bf y}^{3N} 
\delta(E-H)y_{i\alpha}  y_{j\mu}\frac{\partial^2  u_k}{\partial  y_{i\beta}\partial  y_{j\nu}}, \label{term0m}
\eea

Performing integration by parts as in (\ref{term}) and omitting the integrated term we obtain
\bea
&&-\frac{1}{\Omega}\int d {\bf p}^{3N}d {\bf y}^{3N}\frac{\partial  u_k}{\partial  y_{j\nu}}\frac{\partial}{\partial y_{i\beta} }
\big(\delta(E-H)y_{i\alpha}  y_{j\mu} \big) \non \\
&=&-\frac{1}{\Omega}\int d {\bf p}^{3N}d {\bf y}^{3N}\frac{\partial  u_k}{\partial  y_{j\nu}}
\bigg(\frac{\partial \delta(E-H)}{\partial  y_{i\beta}}y_{i\alpha}  y_{j\mu} \non \\&&+\delta(E-H)(\delta_{\alpha\beta}y_{j\mu} 
+\delta_{ij}\delta_{\beta\mu}y_{i\alpha})\bigg). \label{term4}
\eea
The key step now is to show that
\bea
&&-\frac{1}{\Omega}\int d {\bf p}^{3N}d {\bf y}^{3N}
\frac{\partial  u_k}{\partial  y_{j\nu}}
\frac{\partial \delta(E-H)}{\partial  y_{i\beta}}y_{i\alpha}  y_{j\mu} \non \\
&&=\frac{1}{k_BT}\left\bra \frac{\partial  u_k}{\partial  y_{j\nu}}\frac{\partial  U}{\partial  y_{i\beta}}y_{i\alpha}  y_{j\mu}\right\ket, \label{c4}
\eea
where the brackets now denote averaging over the microcanonical ensemble.
Here we follow the approach used in Ref. \cite{Zubarev}.
Let us denote $(\partial  u_k/\partial  y_{j\nu})
y_{i\alpha}  y_{j\mu}$ in the first line of (\ref{c4}) by $G$. We have
\begin{widetext}
\bea
 &&-\frac{1}{\Omega(E)}\int  d{\bf p}^{3N}d{\bf y}^{3N}
 \frac{\partial \delta(E-H)}{\partial y_{i\beta}} G  
 =\frac{1}{\Omega(E)}\int d{\bf p}^{3N}d{\bf y}^{3N}\frac{\partial \delta(E-H)}{\partial E}\frac{\partial U}{\partial y_{i\beta}}G  \non \\
 &&=\frac{1}{\Omega(E)}\frac{\partial}{\partial E}\int d{\bf p}^{3N}d{\bf y}^{3N} \frac{\delta(E-H)}{\Omega(E)}\Omega(E)\frac{\partial U}{\partial y_{i \beta}}G 
 =\int d{\bf p}^{3N}d{\bf y}^{3N} \frac{\delta(E-H)}{\Omega(E)}\left(\frac{\partial\ln\Omega(E)}{\partial E}\right)\frac{\partial U}{\partial y_{i \beta}}G \non \\ 
 &&+\frac{\partial}{\partial E}\int d{\bf p}^{3N}d{\bf y}^{3N}\frac{\delta(E-H)}{\Omega(E)}\frac{\partial U}{\partial y_{i \beta}}G
 =\frac{1}{k_BT}\left\bra \frac{\partial U}{\partial y_{i \beta}}G\right\ket
 +\frac{\partial}{\partial E}\left\bra \frac{\partial U}{\partial y_{i \beta}}G\right\ket. \label{c1}
 \eea
 \end{widetext}
 Here the second expression follows because $\delta(E-H)$ depends on $y_{i\beta}$
 only through the potential energy $U$; the third expression is obtained by moving the 
 derivative with respect to $E$ in front of the integral and inserting $\Omega(E)/\Omega(E)=1$ in the integrand; the fourth and fifth expressions follow because  
 \bea
 &&\frac{\partial}{\partial E}\left(\frac{\delta(E-H)\Omega(E)}{\Omega(E)}\right)=\frac{\delta(E-H)}{\Omega(E)}\frac{\partial \Omega(E)}{\partial E}\non \\&&+\Omega(E)\frac{\partial}{\partial E}\left(\frac{\delta(E-H)}{\Omega(E)}\right);
 \eea
 and the last equality in Eq. (\ref{c1}) is obtained using the fact that 
 $\ln \Omega(E)=S(E)/k_B$ and ${(\partial S/\partial E)}_V=1/T$. If $\bra(\partial U/\partial y_{i \beta})G \ket$ in (\ref{c1}) is finite then the last term in (\ref{c1})
vanishes in the thermodynamic limit because it represents a derivative of $\bra(\partial U/\partial y_{i \beta})G \ket$ with respect to an
extensive variable. This proves the relationship (\ref{c4}). The rest of the proof follows the same steps as in Appendix \ref{Ap2}.
 
%


\end{document}